\pdfoutput=1
\documentclass{article}

\PassOptionsToPackage{numbers, compress}{natbib}
\usepackage[final]{neurips_2022_ml4ps}
\usepackage[utf8]{inputenc} % allow utf-8 input
\usepackage[T1]{fontenc}    % use 8-bit T1 fonts
\usepackage[hidelinks]{hyperref}       % hyperlinks
\usepackage{url}            % simple URL typesetting
\usepackage{booktabs}       % professional-quality tables
\usepackage{amsfonts}       % blackboard math symbols
\usepackage{nicefrac}       % compact symbols for 1/2, etc.
\usepackage{microtype}      % microtypography
\usepackage{xcolor}         % colors
\usepackage{graphicx}
\usepackage{amsthm, amssymb}
\usepackage[symbol]{footmisc}

\usepackage{amsmath}

\renewcommand{\vec}[1]{\boldsymbol{#1}}

\usepackage{setspace}



\title{Statistical Inference for Coadded Astronomical Images}

\author{%
  Mallory Wang\thanks{These authors contributed equally.} \\
  Department of Statistics\\
  University of Michigan\\
  \texttt{wmallory@umich.edu} \\
  \And
  Ismael Mendoza$^*$ \\
  Department of Physics\\
  University of Michigan\\
  \texttt{imendoza@umich.edu} \\
  \AND
  Cheng Wang\\
  Department of Statistics\\
  University of Michigan\\
  \texttt{wangchv@umich.edu} \\
  \And
  Camille Avestruz \\
  Department of Physics \\
  University of Michigan \\
  \texttt{cavestru@umich.edu} \\
  \And
  Jeffrey Regier\\
  Department of Statistics\\
  University of Michigan\\
  \texttt{regier@umich.edu}
}

\begin{document}

\maketitle

\begin{abstract}
Coadded astronomical images are created by stacking multiple single-exposure images. Because coadded images are smaller in terms of data size than the single-exposure images they summarize, loading and processing them is less computationally expensive. However, image coaddition introduces additional dependence among pixels, which complicates principled statistical analysis of them.
We present a principled Bayesian approach for performing light source parameter inference with coadded astronomical images. Our method implicitly marginalizes over the single-exposure pixel intensities that contribute to the coadded images, giving it the computational efficiency necessary to scale to next-generation astronomical surveys. As a proof of concept, we show that our method for estimating the locations and fluxes of stars using simulated coadds outperforms a method trained on single-exposure images.
\end{abstract}

%%%%%%%%%%%%%%%%%%%%%%%%%%%%%%%%%%%%%%%%%%%%%%%%
\vspace{-7pt}\section{Introduction}\vspace{-7pt}
%%%%%%%%%%%%%%%%%%%%%%%%%%%%%%%%%%%%%%%%%%%%%%%%

% Current era of large surveys, what co-addition of images does
The next generation of astronomical surveys, including the Rubin Observatory's Legacy Survey of Space and Time (LSST), will produce massive quantities of image data. Measurements of large structures in our universe based on this data will allow us to constrain cosmological parameters to unprecedented levels of precision \citep{abell2009lsst}. These analyses rely on combining multiple \textit{single-exposure} astronomical images into a single \textit{coadded} image by linearly combining stacked pixel intensities. Relative to the single-exposure images on which they are based, coadds are nearly three orders of magnitude smaller in terms of data size and have greater image depth \citep{annis2014sloan}. As a result, coadded images are much easier to analyze: they are faster to load and have higher signal-to-noise ratios.

% Statistical inference on co-adds
However, coadded images are more difficult to interpret in a statistically principled way than single-exposure images. The single-exposure images contributing to a coadd vary in point spread function (PSF) and require resampling to be aligned to a common grid \citep{mohr2012dark, bosch2019overview}. Resampling and PSF homogenization create additional covariance between nearby pixels, complicating statistical inference on coadds; whereas in the raw images, pixel intensities are well modeled as independent conditional on the latent properties of nearby stars, galaxies, and the background, in coadds they are not. 

% BLISS and how it breaks down for coadds
Bayesian Light Source Separator (BLISS) is a scalable probabilistic method for detecting and cataloging astronomical objects using single-exposure images \citep{hansen2022scalable}. In the current work, we extend the BLISS model to create a fully generative model of coadded images (Section 2). The true coaddition process is deterministic and well understood, so modeling it requires few assumptions. In single-exposure images, BLISS models pixel intensities as conditionally independent given the latent properties of imaged light sources; each pixel has an observed pixel intensity sampled from a Poisson distribution with the underlying intensity, unique to that pixel, as its rate.  
This approach breaks down for coadds, in which pixel intensities are not independent given the latent properties of imaged sources. To account for coadds, we add a unique latent variable for each pixel in a raw image that contributes to the observed coadd.  

% FAVI
We leverage forward amortized variational inference (FAVI) \citep{ambrogioni2019forward}, a computationally efficient approach that implicitly marginalizes over the new latent variables corresponding to the pixel values from single-exposure observations (Section 3). These new latent variables are nuisance variables; we wish to infer only the properties of imaged light sources. FAVI effectively ignores large numbers of nuisance latent variables in its optimization routine. This approach is thus a computationally efficient way of performing principled Bayesian inference on coadded images.

We compare results of four trained BLISS encoders with varying numbers of coadded single-exposures. We find that our method applied to coadds outperform the method applied to single-exposures, in terms of detection and flux metrics (Section 4). Our results illustrate the feasibility of performing principled end-to-end Bayesian inference on image coadds for downstream analyses. 
The source code for these studies is available at \url{https://github.com/prob-ml/bliss}.

%%%%%%%%%%%%%%%%%%%%%%%%%%%%%%%%%%%%%%
\vspace{-7pt}\section{Statistical Model}\vspace{-7pt} \label{sec:dataset}
%%%%%%%%%%%%%%%%%%%%%%%%%%%%%%%%%%%%%%
% High-level single-exposure generation
We first describe the generative model for single-exposure images. We generate single-exposures of size $90 \times 90$ following \citep{liu2021variational}. We generate simulated images of stars using a constant (but realistic) PSF, denoted $\Pi$, from SDSS~\citep{lupton2005sdss}. We characterize stars with two parameters: location $\ell$ and flux $f$. Our single-exposures model an SDSS r-band exposure with only stars. 

% Prior
We sample star parameters from the prior as follows. First, we sample the number of stars $S$ in the image from a Poisson distribution with mean rate $\lambda = 3$. We uniformly sample independent centroids for each source $s = 1, 2,..., S$ within the central $40 \times 40$ pixel square of the image. Finally, the flux follows a truncated power law distribution: $f_{s} \sim \text{Pareto}(f_{\rm min}, f_{\rm max}, \alpha)$, where $f_{\rm min} = 622$ counts, $f_{\rm max} = 10^{6}$ counts, and $\alpha = 0.47$. We denote the catalog $z = \{ \ell_{s}, f_{s} \}_{s=1}^S$ as the collection of the parameters of all the stars in that image. 

\begin{figure}[ht]
    \centering
    \includegraphics[width=\textwidth]{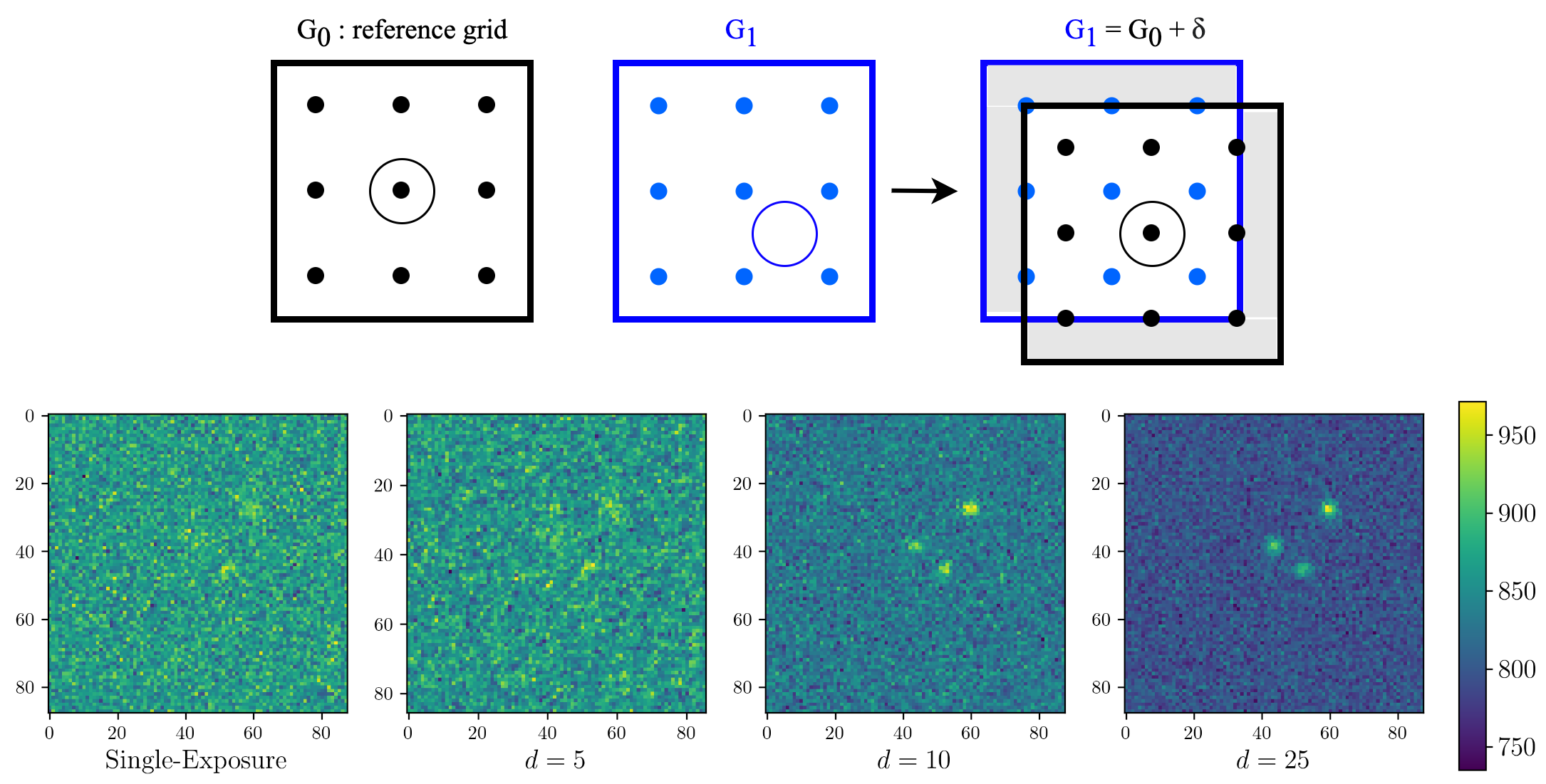} 
    \caption{
        Top: Cartoon outlining grid interpolation for alignment. 
        Bottom: Comparison of a single-exposure (leftmost panel) with coadd images comprised by an increasing number of single-exposures $d=5, 10, 25$. All images contain stars with the same locations and fluxes and have dimensions $88\times88$ pixels. For more details on the generation process see Section~\ref{sec:dataset}.
    }
    \label{fig:cartoon}
\end{figure}

% Alignment process and single exposure
To generate a coadd $y$, we first need to create the set of $d$ aligned single-exposures $\vec{x} = \{x^{(i)}\}_{i=1}^{d}$ that are stacked for a given coadd. To create this set, we repeat the single-exposure procedure $d$ times with the same catalog $z$, but we vary (\textit{dither}) the location of every star in a given single-exposure by the same random sub-pixel shift between $-0.5$ and $0.5$ in both horizontal and vertical directions. We independently sample this location noise for each exposure. 
We then \textit{align} the $d$ single-exposures using a reference grid $G_0$, which corresponds to the not-dithered reference single-exposure $x_0$. We bilinearly interpolate the dithered single-exposures at the grid points $G_0$, then crop the border of all exposures, resulting in $88 \times 88$ images. The result is the set of aligned single-exposures $\vec{x} = \{x^{(i)}\}_{i=1}^{d}$ all evaluated on a common reference grid $G_0$. The top panel of Figure~\ref{fig:cartoon} illustrates our alignment and interpolation procedure. 
Note, the interpolation step is a weighted average between adjacent pixels given the sub-pixel dither. {\it Critically, the alignment procedure induces a correlation between adjacent pixels and has a smoothing effect: each aligned exposure no longer follows a Poisson noise model.}

% Summary of model
In our model, the coadd is a weighted sum of the single-exposures images: $y = \sum_{i=1}^{d} w'_{i} x^{(i)}$ where $w'_i = w_i / \sum_{j=1}^{d} w_{j}$. We use an inverse variance weighting scheme, which results in an optimal signal-to-noise ratio for detecting objects in the coadd \citep{mandelbaum2022psfs}. Specifically, the weights are $w_{i} = (b + u^{(i)})^{-1}$, where $u^{(i)}$ is the signal of the single exposure $x^{(i)}$ and $b$ is the constant background intensity ($b = 865$ counts). 

The bottom panel of Figure~\ref{fig:cartoon} illustrates an example single-exposure and corresponding coadds of $d=5, 10, 25$ from left to right. By increasing $d$, we can increase the signal-to-noise ratio (SNR) of light sources; objects undetectable in a single-exposure may be more visually obvious in a coadd.

%===================================
\vspace{-7pt}\section{Posterior Inference}\vspace{-7pt}
%===================================
%%% Slightly more technical reason why we use FAVI
We wish to approximate the posterior distribution $p(z \mid y)$ of the catalog $z$ given the observed coadd image $y$. However, the coadd image is potentially made up of many single-exposure images $\vec{x} = \{x^{(i)}\}_{i=1}^{d}$. In principle, a fully Bayesian approach would need to model both the catalog $z$ and the random pixel intensities belonging to all single-exposures $\vec{x}$ given the observed coadd $p(z, \vec{x} \mid y)$, and then marginalize over $\vec{x}$ by performing the integral $\int_{\vec{x}} p(z, \vec{x} \mid y) d\vec{x}$. This integral is computationally intractable given the \textit{very} large number of latent variables that need to be marginalized over (one per pixel of each single-exposure making up the coadd). For our model, common Bayesian inference techniques such as MCMC and traditional variational inference would be extremely computationally expensive to perform because they would require explicitly modeling large numbers of nuisance latent variables. In general, these nuisance latent variables cannot be marginalized analytically. Instead, we use Forward Amortized Variational Inference (FAVI) \citep{ambrogioni2019forward}, a recently developed likelihood-free approach to amortized variational inference that uses the forward Kullback-Leibler (KL) divergence in a joint-contrastive variational loss. 

% How FAVI helps us in the context of coadds
Assuming a flexible enough encoder, the global optima of the FAVI loss is guaranteed to be $p(z \mid y)$ with a FAVI loss that consistently marginalizes $\vec{x}$.  The FAVI loss implicitly marginalizes over $\vec{x}$ simply by excluding these variables from the objective, without having to consider their dependency with the catalog $z$. This allows us to approximate $p(z \mid y)$ directly with a variational distribution $q(z \mid y)$, and optimize choosing $q$ based on the FAVI loss as described in detail next. 

%===================================
\vspace{-7pt}\subsection*{Training and Architecture}\vspace{-7pt}
%===================================
% What BLISS does
Our inference routine reuses BLISS \citep{hansen2022scalable} code, which already employs the FAVI loss to infer light source counts, locations, and fluxes in single-exposure images.  
In \cite{hansen2022scalable}, the input to the encoder neural network consists of single-exposure images, each split into equal-sized \textit{tiles}. The weights of the neural network, $\phi$, parametrize a variational distribution $q_{\phi}( z \mid x )$.  The variational distribution approximates the true marginal posterior, $p(z\mid x)$, on the catalog, $z$, given a single-exposure, $x$. We choose a variational distribution that factorizes over the tiles. For each tile, the variational distribution further factorizes over a categorical distribution for the number of light sources, a logit distribution for the (normalized) locations, and a log-normal distribution for the fluxes. The encoder outputs the parameters in each tile, which are used to compute FAVI loss on each tile and then summed over all tiles of each image. This loss is used to optimize the neural network weights $\phi$. 

% Change for this application/extension
To extend BLISS to coadds, we retrain the BLISS encoder uses the training procedure in \cite{hansen2022scalable, liu2021variational} to approximate the marginal posterior of the true catalog given the coadd with a new variational distribution $q_{\phi}( z \mid y )$. The only modification required is to use coadds as images for training instead of single-exposures. We train four different encoders with different numbers of $d$ coadded single-exposures: 1, 5, 10, 25. The encoder trained with one exposure is just the original BLISS encoder trained on single-exposures with no dither and no interpolation. For each of these encoders, we fitted a variational distribution that outputs counts, locations, and fluxes of stars. 

% Architecture and training specification
The neural network in the encoder consists of a combination of convolutional, batch-norm, and dropout layers using ReLUs as an activation. We use PyTorch \citep{paszke2019pytorch} and the Adam optimizer with a learning rate of $10^{-4}$. For optimization, we use an NVIDIA GeForce RTX 2080 Ti GPU. For both the single-exposure encoder and the coadd encoders we trained for 30 epochs, where each epoch consisted of a fixed set of 10000 images of size $88 \times 88$. Each encoder type was trained on the same dataset five times with different random seeds in order to estimate epistemic uncertainty.

%%%%%%%%%%%%%%%%%%%%%%%%%%%%%%%%%%%%%%
\vspace{-7pt}\section{Validation Metrics and Results}\vspace{-7pt}
%%%%%%%%%%%%%%%%%%%%%%%%%%%%%%%%%%%%%%

\begin{figure}[ht]\vspace{-8pt}
    \centering
    \includegraphics[width=\textwidth]{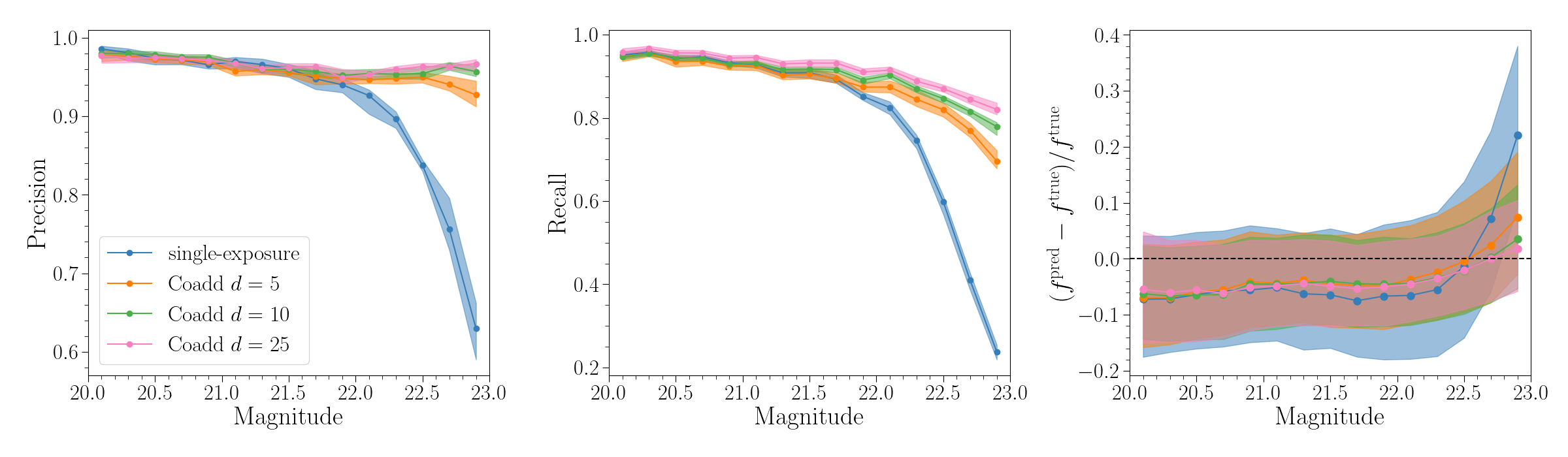}\vspace{-3pt}
    \caption{ 
        Left and center: Precision and recall of source identification, computed by matching predictions with true locations within one pixel. Bands indicate 25--75 percentiles of the bootstrap samples across the five training seeds. 
        Right: Median fractional residual of the predicted flux of each matched light source, with bands indicating 25--75 percentile values of the residuals.
    }
\label{fig:money}
\end{figure}

% Describe Precision/Recall/Flux residual (money) figure
Figure~\ref{fig:money} shows the results of the trained BLISS encoders with increasing numbers of single exposures ($d = 1, 5, 10, 25$), evaluated on test datasets with their respective number of coadded single-exposures. The results we report are based on the mode of the variational distribution for star locations and fluxes.

% Observations of precision/recall
For objects brighter than $21.5$ magnitude,\footnote{As a reference, the brightest stars we considered in our results with magnitude $20$ correspond to an (isolated) SNR of around $45$ on a single exposure and $100$ for the coadd with $d=25$. On the other hand, the dimmest stars with magnitude $23$ have an SNR of around $3$ for the single exposure and $7.5$ on the coadd with $d=25$.} the precision and recall values are similar for the single-image and coadded-image results.  However, as objects become dimmer, the performance of the single-exposure encoder rapidly decreases, with the recall reaching $0.25$ and precision reaching $0.65$ in our dimmest objects of about $23$ magnitude. For the coadd encoders, the precision remains mostly constant for all magnitudes. The recall for coadd encoders starts decreasing past magnitude $21.5$, but stays above $0.70$ for even the dimmest objects. The recall across magnitudes in Figure~\ref{fig:money} follows the expected ordering with higher number of single-exposure in coadds.\footnote{We noticed a saturation in detection performance for coadd encoders with $d > 25$. The difference in recall and precision became statistically insignificant past $d=25$.}

% Observations of flux including uptick due to noise.
Finally, the residual flux median are nearly zero for all magnitudes, but the residual fluxes from the single-exposures have larger scatter than the coadd ones. This suggests higher confidence predictions on flux by the coadd encoders, which makes sense given the higher SNR in their images.
Moreover, the single-exposure encoder predicts a residual median that exceeds $20\%$ in the dimmest magnitude bin. All coadd encoders output an absolute median residual below $10\%$ and have lower scatter than that of the single-exposure. Thus, the coadd method outperforms the single-exposure method in predicting flux.

%%%%%%%%%%%%%%%%%%%%%%%%%%%%%%%%%%%%%%
\vspace{-7pt}\section{Conclusion}\vspace{-7pt}
%%%%%%%%%%%%%%%%%%%%%%%%%%%%%%%%%%%%%%

We present a novel fully Bayesian approach for performing light source parameter inference on coadded images. Our method implicitly marginalizes over the single-exposure pixel intensities that contribute to the coadd and is computationally efficient. We show preliminary results of our model in estimating star locations and fluxes using simulated coadds containing stars. Our method on coadds outperforms that of the single-exposure in both detection and flux metrics. Thus, it has the potential to scale to next-generation astronomical surveys and provide calibrated uncertainties for the measured properties of potentially billions of detected light sources. Future work will be devoted to extending the model to more realistic simulated coadds. Additionally, we will propagate uncertainties to downstream analyses, such as in photometric redshift estimation. 

{\bf Broader Impact: }Our method has potential implications for quantifying uncertainty in the context of classification and detection in noisy images. Given the differences between astronomical images and real-world images, we do not expect our method to be misused for purposes like surveillance. 
Our approach can have a positive impact on the growing environmental cost of computing. Whereas other solutions such as MCMC may require more time and computational power, our approach has the potential to be more efficient in both aspects. 

%%%%%%%%%%%%%%%%%%%%%%%%%%%%%%%%%%%%%%
\begin{ack}
IM and CA acknowledge support from DOE grant DE-SC009193.
\end{ack}
%%%%%%%%%%%%%%%%%%%%%%%%%%%%%%%%%%%%%%

%%%%%%%%%%%%%%%%%%%%%%%%%%%%%%%%%%%%%%%%%%%%%%%%%%%%%%%%%%%%%%%%%%%%%%%%%%%%
\bibliography{main}
%%%%%%%%%%%%%%%%%%%%%%%%%%%%%%%%%%%%%%%%%%%%%
\end{document}